\documentstyle[12pt]{article}
\renewcommand{\thefootnote}{\fnsymbol{footnote}}
\begin{document}
\begin{flushright}
Frankfurt preprint UFTP--436 \\
Duke preprint DUKE--TH--97--139\\
Columbia preprint CU--TP--820
\end{flushright}
\vspace*{1cm}
\setcounter{footnote}{1}
\begin{center}
{\Large\bf Kaon Interferometry as Signal for }\\
{\Large\bf the QCD Phase Transition at RHIC
\footnote{Supported by DFG, BMBF, GSI, and DOE.}}
\\[1cm]
Stefan Bernard $^1$, Dirk H.\ Rischke $^2$, Joachim A.\ Maruhn $^1$, 
Walter Greiner $^1$ \\
\vspace*{1cm}
{\small $^1$ Institut f\"ur Theoretische Physik der 
J.W.\ Goethe--Universit\"at} \\
{\small Robert--Mayer--Str.\ 10, D--60054 Frankfurt/M., Germany}\\
\vspace*{1cm}
{\small $^2$ Department of Physics, Duke University} \\
{\small Durham, NC 27708--0305, U.S.A.}
\\ ~~ \\ ~~ \\
{\large March 1997}
\\[1cm]
\end{center}
\begin{abstract}
Pion and kaon correlations in relativistic nuclear
collisions are studied 
in the framework of boost-invariant, cylindrically symmetric
hydrodynamics. It is investigated how the inverse widths, $R_{\rm out}$,
$R_{\rm side}$, of the two--particle
correlation functions in out-- and side--direction depend
on the average transverse momentum $K_{\bot}$ of the 
particle pair, the initial energy density $\epsilon_0$, and
the equation of state of the system.
The QCD transition leads to a time delay in the expansion of
the system and consequently to an enhancement of the ratio 
$R_{\rm out}/ R_{\rm side}$.
This time--delay signal is found to be particularly strong for large average
transverse momenta $K_\bot \sim 1$ GeV and initial energy densities 
accessible at RHIC, $\epsilon_0 \sim 10 - 20\, {\rm GeV\, fm}^{-3}$. 
Neutral kaon pair correlation functions, which are not influenced 
by final state Coulomb effects and less contaminated by resonance decays 
than pion correlation functions, seem to be
the ideal tool to detect this collective time--delay signature of
the QCD transition.
\end{abstract}
\renewcommand{\thefootnote}{\arabic{footnote}}
\setcounter{footnote}{0}
\newpage
\section{Introduction}
Lattice studies of quantum chromodynamics (QCD) exhibit
a chiral symmetry restoring, deconfining transition around a
temperature $T_c \sim 160$ MeV (for (net) baryon-free matter) \cite{Lattice}.
The only possibility to study this transition and the properties of the
high-temperature phase of nuclear matter, the so-called quark--gluon plasma
(QGP) \cite{muller}, under laboratory conditions is via 
relativistic heavy-ion collision experiments \cite{harris}.

Promising signals for the detection of the QGP emerge from the
influence of the QCD equation of state (EoS) on the
collective dynamical evolution of the many-particle system.
Such signals have to be studied via relativistic hydrodynamics
\cite{strottman}, since
that is the only dynamical model which provides a direct link between 
collective motion and the EoS of strongly interacting matter.
For instance, it was shown in \cite{Shuryak,RisBerM95,RisGy96} that
the QCD transition softens the EoS in the transition
region and therefore reduces the tendency of matter to expand.
This has two important consequences.
On one hand, the expansion of the system is delayed and its lifetime
is considerably prolonged \cite{Shuryak,RisBerM95,RisGy96}.
On the other hand it was shown in \cite{Csernai,RisPuretal} that the
reduced expansion tendency of matter leads to a reduction of the 
transverse directed flow in semi-peripheral heavy-ion collisions.

In \cite{RisGySig} the hydrodynamic flow pattern
was studied for spherically symmetric (so-called ``fireball'') 
expansion as well as
for cylindrically symmetric transverse expansion with
longitudinally boost-invariant initial conditions (so-called ``Bjorken
cylinder'' expansion), both as
a function of the initial energy density $\epsilon_0$ and a
possible finite width $\Delta T$ of the QCD transition region.
The QCD transition was shown to considerably prolong the lifetime of the
expanding system as compared to
the expansion of an (ultrarelativistic) ideal gas, provided the
initial energy density lies in a certain range of values.
This range corresponds to AGS initial energy densities
for the fireball geometry, and RHIC initial energy densities for the
Bjorken cylinder geometry (for realistic values of the initial thermalization
time scale $\tau_0$). Motivated by Pratt's and Bertsch's
conjecture \cite{Pratt,Bertsch}, it was shown in \cite{RisGySig}, 
that for both geometries the prolongation
of the lifetime could be experimentally observable via the ratio
$R_{\rm out}/ R_{\rm side}$ of the inverse widths of the
two--particle correlation functions in out-- and
side--direction, since this ratio follows the behaviour of
the lifetimes rather closely.

In this paper we extend the investigations of \cite{RisGySig} for
the Bjorken cylinder geometry in two ways.
First, we study two--kaon in addition to two--pion
interferometry. While pion correlation functions are contaminated to a large
extent by resonance decays, one expects these to be of minor influence
in the case of kaons \cite{resonance}. In addition, neutral kaon 
correlation functions are, in contrast to the pionic ones, not distorted by 
final state Coulomb effects 
which introduce an additional uncertainty in experimental data.

Second, we now present a systematic study of 
$R_{\rm out}$, $R_{\rm side}$, and their ratio 
as a function of initial energy density $\epsilon_0$ {\em and\/} of
the mean transverse pair momentum $K_{\bot}$ at midrapidity.
As in \cite{RisGySig}, we take an EoS with a first order phase transition 
($\Delta T=0$) as well as an EoS with a finite width $\Delta T=0.1\, T_c$ 
(as an upper limit deduced from QCD lattice calculations \cite{Lattice}), and
compare the enhancement of $R_{\rm out}/ R_{\rm side}$
relative to the ideal gas case.

We find that the excitation function of $R_{\rm out}/ R_{\rm side}$
for kaons 
again mirrors closely that of the lifetime of the system, and has a maximum
around the same initial energy density $\epsilon_0$ as for pions. Due to
kinematical reasons the ratio is, however, 
somewhat smaller for kaons with small
average transverse pair momentum $K_{\bot} < m_K = 497$ MeV 
as compared to the pion case. On the other hand, for high
$K_{\bot} \sim 1$ GeV, it is nearly of the same order as that for pions. 

We also study separately the dependence of $R_{\rm out}$
and $R_{\rm side}$ on $\epsilon_0$ and $K_\bot$. As expected, these
quantities in general decrease with increasing $K_\bot$ at fixed
$\epsilon_0$. An important exception is the situation around
values of $\epsilon_0 \sim 10 - 20 \, {\rm GeV\, fm}^{-3}$, 
where the time--delay signal
is most pronounced. For these initial energy densities and in the
case of a first order transition, $\Delta T=0$,
we find that $R_{\rm out}$ 
actually increases as a function of $K_\bot$, both for pions and kaons.
We analyze this behaviour in detail studying the emission function
and its variances \cite{heinz}.

The outline of the paper is as follows. In Section 2 we briefly discuss
the hydrodynamic model, the EoS used, and
the basic formulae to derive single inclusive particle
spectra and two--particle correlation functions from
the hydrodynamic solution.
In Section 3 we present our results. A summary and conclusions are given 
in Section 4. An appendix contains technical details on the
calculation of the variances of the emission function.
Natural units $\hbar = c = k_B = 1$ are used throughout 
this work.

\section{Hydrodynamics and Par\-ti\-cle Spec\-tra}

\subsection{The hydrodynamic model}

Hydrodynamics is equivalent to local conservation of energy and momentum,
\begin{equation} \label{dTmunu}
\partial_{\mu} T^{\mu \nu} = 0~,
\end{equation}
plus additional conservation equations for the (net) 4--current
of any conserved charge in the system. The latter will not be considered
here, since our focus of interest, 
the central (rapidity) region of ultrarelativistic heavy-ion collisions,
is expected to be essentially free of (net) charges due
to the limited nuclear stopping power.

In order to solve the relativistic hydrodynamic equations 
(\ref{dTmunu}) we make
the assumption of local thermodynamic equilibrium,
the so-called ``ideal fluid approximation''.
This approximation, although commonly employed throughout the literature,
is rather restrictive, since the fluid is certainly out of
thermodynamic equilibrium in the early and late stages of
a heavy-ion collision. On the other hand, a commonly accepted, 
causal theory of dissipative relativistic hydrodynamics, which
would account for non-equilibrium phenomena,
does not yet exist \cite{strottman}. For the moment 
we have to rely on the assumption that (local) thermalization does 
indeed occur after the initial pre-equilibrium stage which is
dominated by (hard) parton--parton scattering processes (for
collisions at RHIC energies). Arguments in favour of this 
assumption are made in \cite{hotglue}. We then describe
the system's evolution from this point on via ideal fluid-dynamics, until
dissipative effects in the final stages prior to decoupling become
large and the assumption of local thermodynamical equilibrium again breaks
down.

In the ideal fluid approximation,
the energy--mo\-men\-tum tensor reads \cite{LanLif}
\begin{equation} \label{Tmunu}
T^{\mu \nu} = (\epsilon + p)\, u^{\mu} u^{\nu} - p\,
g^{\mu \nu}~.
\end{equation}
Here, $\epsilon,\, p$ are energy density and pressure in the rest 
frame of the fluid, while
$u^{\mu} = \gamma\, (1,{\mbox{\boldmath $\beta$}})$ and
$g^{\mu \nu} = {\rm diag}(+,-,-,-)$
are the 4--velocity of the fluid and the metric tensor, respectively
($\mbox{\boldmath $\beta$}$ is the 3--velocity,
$\gamma \equiv (1-{\mbox{\boldmath $\beta$}}^2)^{-1/2}$ is the gamma factor).

Our model for the dynamics of central
ultrarelativistic heavy-ion collisions is
the so-called ``Bjorken cylinder'' expansion \cite{Bjorken},
i.e., it assumes cylindrical symmetry perpendicular to the beam axis
(taken to be the $z-$axis throughout this work)
and a scaling solution for the longitudinal expansion
($\beta^z = z/t$). 
Note that due to the details of the mini-jet production mechanism,
these assumptions about the symmetries of the system are 
not valid per single event \cite{MGDHRBZ}. On the other hand, they seem
quite reasonable on an event-average basis.

A consequence of longitudinal scaling is
boost--invariance along the beam axis. It is therefore sufficient
to solve the system of four hydrodynamic equations (\ref{dTmunu})
at $z=0$, where it reduces to
\begin{eqnarray} \label{eom2a}
\partial_t\, E + \partial_r\, [(E+p){\beta}] & = &
-\, \left( \frac{\beta}{r} + \frac{1}{t}
\right) (E +p)\, \, , \\
\partial_t\, M + \partial_r\, (M{\beta}+p) & = &
-\, \left( \frac{\beta}{r} + \frac{1}{t} \right) M\, \, . \label{eom2b}
\end{eqnarray}
Here, $\beta$ is the radial component of the transverse velocity, and we used
the definitions $E \equiv T^{00}, \, M \equiv T^{0r}$
($r$ as index for the radial component of the energy--momentum tensor).
In order to solve the (effectively (1+1)--dimensional) system 
(\ref{eom2a},\ref{eom2b}), we apply the same scheme used
in \cite{RisGySig}, i.e., we employ Sod's operator
splitting method \cite{Sod} to account for the source terms
on the right-hand side, and the numerically well-tested
relativistic HLLE algorithm \cite{RisBerM95,Schneider} for the
one--dimensional hydrodynamic transport on the left-hand side
of (\ref{eom2a},\ref{eom2b}). The HLLE scheme
was shown to yield excellent results in particular for equations 
of state with phase transitions \cite{RisBerM95,RisPurM95}.

\subsection{The equation of state}

As mentioned in the introduction, the advantage of 
hydrodynamics to model the dynamics of
relativistic heavy-ion collisions is
that this is the only approach which directly relates 
observable collective flow phenomena
to the EoS of hot and dense nuclear matter.
Formally, specifying an EoS closes the coupled system
of hydrodynamic equations. In the particular case
of vanishing conserved (net) charges considered here, the system
(\ref{eom2a},\ref{eom2b}) is closed by an EoS of the form
$p(\epsilon)$. In the ideal fluid approximation the 
emerging flow pattern is completely determined by the
choice of this EoS and the initial conditions.

In the present work we study hydrodynamic expansion
for three different equations of state.
The simplest one is the EoS of an ultrarelativistic ideal gas,
\begin{equation} \label{EoS1}
  p(\epsilon) = c_s^2\, \epsilon~,
\end{equation}
where $c_s = 1/ {\sqrt{3}}$ is the velocity of sound.
This is commonly taken to be a reasonable first approximation
for a hadronic matter EoS in the
central region of an ultrarelativistic heavy-ion collision,
since here the system is hot, $T > m_\pi$, and thus dominated by pions, 
which are the lightest hadrons. The EoS (\ref{EoS1}) 
has a constant pressure gradient
${\rm d}p/{\rm d}{\epsilon} \equiv c_s^2=1/3$ for all
energy densities. 

To study the effects of a (phase) transition to QGP on the dynamics
of the system, we use the following parametrization of the 
entropy density as function of temperature \cite{RisGy96,RisGySig,Para},
\begin{equation} \label{EoS23}
\frac{s}{s_c}(T) = \left[\frac{T}{T_c}\right]^3
\left( 1 + \frac{d_{\rm Q}-d_{\rm H}}{d_{\rm Q}+d_{\rm H}} \,
\tanh \left[ \frac{T-T_c}{\Delta T} \right] \right)~.
\end{equation}
Here, $s_c=(d_{\rm Q}+d_{\rm H}) {\pi}^2 T_c^3 /45$
is the entropy density at the critical temperature $T_c$,
and $d_{\rm Q}$ and $d_{\rm H}$ are
the number of (massless) degrees of freedom in the QGP and the hadronic
phase, respectively (we fix $d_{\rm Q}=37$ and $d_{\rm H}=3$
throughout this work, for the effect of other choices on the
dynamics of the system, see \cite{RisGySig}).
The function $p({\epsilon})$ then follows from thermodynamical
relations and is used in the hydrodynamic model in tabular form.
Eq.\ (\ref{EoS23}) accounts for a possible finite width
$\Delta T$ of the transition. Present QCD
lattice data \cite{Lattice} indicate $0 \le \Delta T < 0.1\, T_c$.

In the limit $\Delta T=0$, eq.\ (\ref{EoS23}) exhibits
a first order phase transition between an
ultrarelativistic gas of hadrons described by the EoS
(\ref{EoS1}), and a (net) baryon-free QGP consisting
of gluons and $u$ and $d$ quarks (and antiquarks) 
described by the well-known MIT bag EoS \cite{bag}.
Energy density and pressure in the QGP phase are then
${\epsilon}_{\rm q}=d_{\rm Q} {\pi}^2 T^4/30 + B$,
$p_{\rm q}=(\epsilon_{\rm q}- 4 B)/3$, while the corresponding quantities
in the hadronic phase are
${\epsilon}_{\rm h}=d_{\rm H} {\pi}^2 T^4/30$,
$p_{\rm h}=\epsilon_{\rm h}/3$.
$B=\frac{1}{2} \left[(d_{\rm Q}-d_{\rm H})/(d_{\rm Q}+d_{\rm H})\right] 
T_c s_c$
is the MIT bag constant, if one measures energy densities in units
of the critical enthalpy density $w_c \equiv \epsilon_c + p_c = T_c s_c$
($p_c$ is the pressure at $T_c$; for our choice 
$d_{\rm Q}/d_{\rm H}=37/3$ and $T_c=160$ MeV,
$T_c s_c \simeq 0.75\,{\rm GeV\, fm}^{-3}$ in physical units).
Besides this first order transition with $\Delta T=0$, we also consider
a transition with a finite width $\Delta T=0.1\, T_c$
to cover present uncertainties in the QCD EoS.
Note, however, that a finite width would also naturally emerge in any
finite system, even if the transition is of first order 
in the thermodynamic limit \cite{carsten}.

The three different equations of state are displayed in Fig.\ 1. 
Part (a) shows the entropy density divided by $T^3$
and (b) the energy density divided by $T^4$ as functions
of temperature. For the equations of state with a transition, at $T=T_c$
both quantities exhibit the strong increase (or even a
discontinuity) observed in lattice calculations. Part
(c) shows the pressure and (d) the velocity of sound squared,
$c_s^2 \equiv {\rm d}p/{\rm d}{\epsilon}$, as functions of energy
density.
While for the ideal gas EoS (dashed lines) $c_s^2=1/3$ is 
constant for all energy densities,
this is different for the EoS with a
first order phase transition, $\Delta T=0$ (full lines),
and the EoS with a transition of finite width 
$\Delta T=0.1\, T_c$ (dotted lines).
The pressure gradient $c_s^2={\rm d}p/{\rm d}{\epsilon}$ 
vanishes for $\Delta T=0$ as long as matter is
in the mixed phase,
$\epsilon_{\rm H} \equiv \epsilon_{\rm h}(T_c) 
< \epsilon < \epsilon_{\rm q}(T_c) \equiv \epsilon_{\rm Q}$.
For parts of the system, which pass 
through the corresponding region of energy densities,
the tendency to expand is then considerably reduced 
compared to the ideal gas case,
which results in a prolonged lifetime of the system
\cite{RisBerM95,RisGy96,RisGySig,Pratt}. For the transition with finite width
$\Delta T=0.1\, T_c$, the pressure
gradients are finite, but nevertheless smaller than for
the ideal gas. Thus, again the system's lifetime will be
prolonged \cite{RisGy96,RisGySig}.

\subsection{Particle spectra and interferometry}

In a heavy-ion collision, the ideal fluid approximation breaks 
down in parts of the system which become so dilute that 
the collision rates between the particles are smaller
than the (local) expansion rate of the system and thus local
thermodynamic equilibrium can no longer be maintained.
These parts of the system decouple (``freeze out'') from the 
hydrodynamic evolution and, eventually, the corresponding
particles freely stream to the detectors.
In a rigorous sense, on one hand they can no longer be described in the 
hydrodynamical framework, while on the other, the energy and 
momentum carried away by these particles might influence the hydrodynamic
evolution of the rest of the system.

Although attempts have been made \cite{bugaev,jjn}, up to now 
no commonly accepted, consistent description of this so-called ``freeze-out'' 
process exists for the hydrodynamical description of heavy-ion collisions. 
The common way to treat this problem is to solve the 
hydrodynamical equations in the whole forward light cone 
and a posteriori identify those parts of the system as decoupled,
which have cooled below a certain freeze-out temperature $T_f$ or 
whose density has dropped below a freeze-out density $n_f$.
Final particle spectra are then calculated along the
freeze-out hypersurface, $\Sigma(x)$, defined by the condition
$T(x)=T_f$, or $n(x)=n_f$, respectively.

Here we also follow this prescription and employ the
approach of Cooper and Frye \cite{CooFry74} to
calculate the particle spectra
(see e.g.\ \cite{BerMGRis96} for a (1+1)--dimensional
application). In our case of vanishing (net) charges, the particle
density is a function of temperature only, and thus freeze-out at
constant density is equivalent to freeze-out at constant 
temperature. The freeze-out temperature is here taken as a free 
parameter in the range $(0.7-1)\, T_c \simeq (110-160)$ MeV.

The single inclusive momentum distribution of ``frozen-out'' particles,
i.e., particles whose world lines
cross the 3--dimensional hypersurface $\Sigma$
in 4--dimensional space--time is then given by
\cite{CooFry74}
\begin{equation} \label{single}
E\, \frac{{\rm d}N}{{\rm d}^3 {\bf k}} = \frac{d}{(2\pi)^3}
\int_{\Sigma}\, {\rm d}\Sigma \cdot k \,\,
f\left( k \cdot u/T \right)\,\, .
\end{equation}
In our case of pions or kaons, $f(z)=(e^z-1)^{-1}$
is the Bose--Einstein distribution function.
${\rm d}{\Sigma}_{\mu}$ is the normal vector on the (local) hypersurface
element ${\Sigma}^{\mu}$, and the integration runs over
the complete freeze-out hypersurface $\Sigma$. Note that,
while $T$ is per our definition of freeze-out constant along $\Sigma$,
the fluid velocity $u^{\mu}$ is not.

The Cooper--Frye formula (\ref{single}) is strictly valid only along
space-like parts of the hypersurface $\Sigma$, i.e., where the
normal vector ${\rm d} \Sigma_\mu$ is time-like. For time-like parts
of $\Sigma$ (space-like ${\rm d} \Sigma_\mu$), however, 
${\rm d}\Sigma \cdot k$ can in principle become negative,
corresponding to particles which re-enter the fluid and thus do not 
decouple from the system. From the physical point
of view, this ``negative'' number of particles should
not enter the single inclusive spectrum of frozen-out particles
on the left-hand side of
(\ref{single}). Therefore, modifications of the Cooper--Frye
formula have been proposed \cite{bugaev,sinyukov}, which correct for this. 
They shall, however, not be employed here,
because for the hydrodynamic solutions studied in the following we found
the collective flow across the freeze-out surface to be rather strong.
The maximum of the single--particle distribution function $f$
is therefore shifted in a way which renders the number 
of particles with given 4--momentum $k^\mu$ that
actually re-enter the fluid negligible. 
We shall quantify this statement below (cf.\ discussion of Figs.\ 7, 8).

The two--particle correlation function of two identical particles
with momenta ${\bf k}_1,\, {\bf k}_2$ is defined as the ratio
of the two--particle probability
$P({\bf k}_1, {\bf k}_2)$ to the product of the 
single--particle probabilities for uncorrelated particles,
\begin{equation} \label{defC2}
C_2 ({\bf k}_1, {\bf k}_2) = \frac{P({\bf k}_1, {\bf k}_2)}
     {P({\bf k}_1) P({\bf k}_2)}\,\, .
\end{equation}
We follow the method outlined in \cite{RisGySig,Pratt,Sinyukov,Marburg}
to compute (\ref{defC2}). Introducing
the average 4--momentum of the particle pair
$K^{\mu} = (k_1^{\mu}+k_2^{\mu})/2$, and
the relative 4--momentum $q^{\mu}=k_1^{\mu} - k_2^{\mu}$,
and assuming that the particle
source is chaotic and sufficiently large,
the two--particle correlation function for bosons
can be written as \cite{RisGySig,Marburg}
\begin{equation} \label{explC2}
C_2 ({\bf k}_1, {\bf k}_2) = 1 + \frac{\left| \frac{d}{(2\pi)^3}
\int_{\Sigma}\, {\rm d}{\Sigma} \cdot K\,\,
\exp\, [i\, q \cdot {\Sigma} ] \,\,
f\left( K \cdot u/T\right)\right|^2}
{ E_1 \, [{\rm d}N/ {\rm d}^3 {\bf k}_1] \,\,\, E_2\,
[ {\rm d}N/{\rm d}^3 {\bf k}_2] }\,\, .
\end{equation}
For the Bjorken cylinder geometry, the number of independent
variables can be reduced due to rotational symmetry
around the beam axis. As in \cite{RisGySig}, we focus on the correlation
functions of particles emitted at midrapidity $(K^z=q^z=0)$. 
We choose the coordinate system such that
${\bf K}=(K_\bot,0,0)$, ${\bf q}_{\rm out}=(q_{\rm out}, 0, 0)$ 
for the component of ${\bf q}$ parallel to ${\bf K}$, and
${\bf q}_{\rm side}=(0, q_{\rm side}, 0)$ for the component of
${\bf q}$ perpendicular to ${\bf K}$. With these assumptions,
$C_2(K, q_{\rm out}, q_{\rm side})$
becomes a function of three independent variables only.

The so-called out--and side--correlation functions are now defined as
\begin{eqnarray}
C_{2,{\rm out}} (q_{\rm out}) & \equiv & C_2(K_{\bot}, q_{\rm out}, 0) \, , \\
C_{2,{\rm side}} (q_{\rm side})& \equiv & C_2(K_{\bot}, 0, q_{\rm side}) \,.
\end{eqnarray}
We then define the inverse width of the correlation functions
in out-- and side--direction by
\begin{equation} \label{Routside}
R_{\rm out} \equiv \frac{1}{ q_{\rm out}^*} \quad , \quad
R_{\rm side} \equiv \frac{1}{ q_{\rm side}^*} \quad,
\end{equation}
where $q_{\rm out}^*$ is determined by  
$C_{2,{\rm out}}(q_{\rm out}^*)=1.5$, and $q_{\rm side}^*$
analogously.
The inverse widths (\ref{Routside}) are a qualitative measure for the
duration of particle emission and the 
transverse size of the source. Their quantitative values,
however, depend on the initial size of the system, $R_0$,
on the transverse momentum $K_\bot$, as well as on
the details of the hydrodynamic flow pattern at freeze-out
\cite{Pratt,heinz,Marburg}.
As argued in \cite{RisGySig}, these dependences
should in principle largely cancel out if one
considers the {\em ratio\/} of the inverse widths, 
$R_{\rm out}/ R_{\rm side}$.
In \cite{RisGySig} it was also shown that, for the spherical
expansion as well as for the Bjorken cylinder expansion, this
ratio is a good {\em qualitative\/} 
measure for the lifetime of the system.
The explicit form of (\ref{explC2}) for the Bjorken cylinder
geometry was given in Appendix B of \cite{RisGySig} and shall
not be repeated here.

For the side--correlation function, the single--particle spectra
in the denominator of (\ref{explC2}) are equal (because of
rotational symmetry, the single--particle spectrum can only
depend on the modulus of the transverse momentum, which
is equal for the two particles, ${\bf k}_1 = (K_\bot,q_{\rm side}/2,0),\,
{\bf k}_2 = (K_\bot,-q_{\rm side}/2,0)$ for pairs at midrapidity), 
while for the
out--correlation function, they are approximately equal as long as
the modulus of the relative momentum, $q_{\rm out}$, of the particle pair is 
small compared to $K_\bot$ (note that
${\bf k}_1 = (K_\bot+q_{\rm out}/2,0,0),\,
{\bf k}_2 = (K_\bot-q_{\rm out}/2,0,0)$ for pairs at midrapidity).
Furthermore, in this case $K^0\equiv (k^0_1 + k^0_2)/2 \simeq
\sqrt{K_\bot^2 + m^2} \equiv E_K$, 
and the correlation function can be approximated as \cite{heinz}
\begin{equation} \label{apprC2}
C_2({\bf K},{\bf q}) \simeq 1 + \left| \frac{ \int {\rm d}^4 x\, 
\exp[iq \cdot x]\, S(x,K)}{\int {\rm d}^4 x\, S(x,K)} \right|^2
\equiv 1 + | \langle \exp[i q \cdot x] \rangle |^2\,\,\, ,
\end{equation}
where
\begin{equation} \label{emiss}
S(x,K) \equiv \frac{d}{(2 \pi)^3}\,
\int_\Sigma {\rm d} \Sigma \cdot K \, f(K \cdot u/T)\,
\delta^{(4)}(x-\Sigma) 
\end{equation}
is the so-called {\em emission function\/} of the source,
and the right-hand side of (\ref{apprC2}) defines the average
$\langle\, \cdot \,\rangle$. In the following we shall make the
Boltzmann approximation for the single--particle 
distribution function, $f(z) \simeq e^{-z}$,
which usually holds to good accuracy, especially
for large average pair momenta $K_\bot$.

In the case that the emission function is Gaussian in space--time,
the correlation function can be written in the form (again, we
focus on midrapidity pairs, $K^z = q^z = 0$) \cite{heinz}
\begin{equation} \label{gaussC2}
C_2(K_\bot, q_{\rm out}, q_{\rm side}) \simeq  
1 + \exp \left[ -\hat{R}_{\rm side}^2 \, q_{\rm side}^2
- \hat{R}_{\rm out}^2\, q_{\rm out}^2 \right]\,\, ,
\end{equation}
where the ($K_\bot$--dependent) so-called ``side--''
and ``outwards radii'' $\hat{R}_{\rm side}, \,
\hat{R}_{\rm out}$ of the correlation function
are related to the variances of the emission function via \cite{heinz}
\begin{eqnarray}
\hat{R}_{\rm out}^2 & \equiv & \langle \tilde{x}^2 \rangle  - 2\, \beta_K \,
\langle\, \tilde{x} \tilde{t}\, \rangle + \beta_K^2\, \langle \tilde{t}^2
\rangle \label{rout} \,\,\, , \\
\hat{R}_{\rm side}^2 & \equiv & \langle \tilde{y}^2 \rangle\,\,\, .
\label{rside}
\end{eqnarray}
Here $\beta_K = K_\bot/E_K$ is the (approximate)
transverse velocity of the particle pair and
$\tilde{x} \equiv x - \langle x \rangle$ ($\tilde{t},\,
\tilde{y}$ are defined analogously).
Note that, in the case of a Gaussian source leading to the
correlation function (\ref{gaussC2}), the inverse widths defined
in (\ref{Routside}) are related to the radii (\ref{rout},\ref{rside})
via $R_{\rm out,\, side} \simeq 1.2\, \hat{R}_{\rm out,\, side}$.
As will be seen in the next section, this formula holds to
astonishingly good approximation in the cases considered here, even
when the emission functions do not resemble Gaussians.
The eqs.\ (\ref{rout},\ref{rside}) therefore allow for a reasonable
qua\-li\-ta\-tive and quantitative understanding of the shape 
of the source or the emission function, respectively.
The explicit formulae for the variances required to 
calculate $\hat{R}_{\rm out,\, side}$ for
the Bjorken cylinder expansion are given in the appendix.

We finally note that other parametrizations of the correlation
function, such as the Yano--Koonin--Podgoretskii approach \cite{YKP},
may be more advantageous than the standard (Pratt--Bertsch) 
parametrization from the point of view of obtaining a better 
{\em quantitative\/} estimate of the variances of the emission function
\cite{heinz}. For studying the {\em qualitative\/} behaviour of the 
proposed time--delay signature, however, the standard 
parametrization proves to be sufficient.

\section{Results}

For the Bjorken cylinder expansion one has to specify
an initial (proper) time $\tau_0$ for the expansion in order to avoid
the divergence at $t=0$ in eqs.\ (\ref{eom2a},\ref{eom2b}).
Physically, this time corresponds to the thermalization time scale
prior to which initial, non-equilibrium parton--parton scattering processes
dominate the dynamical evolution. Only for times $\tau \geq \tau_0$, 
dynamics can be described by means of (ideal) hydrodynamics.
According to the ``hot-glue'' scenario \cite{hotglue},
$\tau_0\simeq 0.5$ fm for central Au+Au--collisions at RHIC.
Since the transverse radius of the hot central zone in such collisions
is of the order $R_0=5$ fm, we fix $\tau_0 = 0.1\, R_0$ in
the following. The dependence of the hydrodynamic 
flow pattern on varying $\tau_0$ as well as the initial energy density 
$\epsilon_0$ and the EoS was discussed in detail in
\cite{RisGySig} (cf.\ Figs.\ 7-9 therein) and shall not be repeated here.

In particular, it was found in \cite{RisGySig} that, for
fixed average pion pair momentum $K_{\bot}=300$ MeV
and for an initial energy density $\epsilon_0= 18.75\,T_c s_c$,
in the case of the QCD transition the
experimentally measurable ratio $R_{\rm out}/R_{\rm side}$
was considerably enhanced over the ideal gas case,
reflecting the delay in the expansion
of the system due to the softening of the EoS in the transition region.
This is once more shown in Fig.\ 2 (thin lines)
for the hydrodynamic expansion with a first order phase transition
($\Delta T=0$, $d_{\rm Q}/d_{\rm H}=37/3$, $m_\pi=138$ MeV).
The three curves correspond to different freeze-out temperatures 
$T_{f}=0.7\,T_{c}$ (full lines),
$T_{f}=0.9\,T_{c}$ (dotted lines), and $T_{f}=T_{c}$ (dashed lines).
In this work we study this {\em time--delay signature of the QCD transition\/}
not only as a function of $\epsilon_0$, but also its dependence on
$K_\bot$, both for pions and kaons and for the three different
equations of state discussed in Subsection 2.2.

Note that all correlation functions in this work are calculated taking the
finite mass of the respective particles into account, but for the fluid
evolution, hadron matter is considered to be an ultrarelativistic ideal gas. 
While this appears inconsistent from a
rigorous point of view, quantitatively we do not expect a finite
pion mass in the EoS to severely affect the hydrodynamical evolution
prior to decoupling, since the freeze-out temperatures considered
here are around or even above $m_\pi$. Kaons in the hadronic
phase are, on the other hand, suppressed by their larger mass,
$m_K > T_c$. Therefore, the effect of these particles on
the hydrodynamic evolution is expected to be small. This view is supported
by the results of \cite{dinesh}. Moreover, small differences in the
hydrodynamic evolution tend to have an even smaller impact on the
correlation functions (which represent in a certain sense averages over
the dynamical evolution) and thus on $R_{\rm out}$ and $R_{\rm side}$.

The thick lines in Fig.\ 2 represent our new results for
kaons ($m_K=497$ MeV, $K_\bot = 300$ MeV). 
For these heavier mesons we also observe a local
maximum in the excitation function, which again mirrors
closely the prolongation of the lifetime of the system
around $\epsilon_0 \simeq 18.75\, T_c s_c$.
Such a behaviour is {\em not\/} seen
in the ideal gas case, where
$R_{\rm out}/R_{\rm side}$ varies rather smoothly (for the sake
of clarity, we do not show results for the ideal gas in Fig.\ 2;
for that case see, for instance, Figs.\ 3, 5 below and Fig.\ 17 (a) in
Ref.\ \cite{RisGySig}).
Finally, note that for $K_\bot = 300$ MeV, the enhancement
is nearly independent of the freeze-out temperature both
for pions as well as for kaons.

For $K_\bot = 300$ MeV the absolute
magnitude of $R_{\rm out}/R_{\rm side}$ is about a factor of 2
smaller for kaons than for pions. The reason
is the higher mass of the kaons compared to pions which
leads, for the comparatively small value of $K_\bot = 300$ MeV, 
to a considerable difference in the velocity of the kaon 
pair, $\beta_K \simeq 0.52$, as compared to a pion pair with the
same $K_\bot$, $\beta_K \simeq 0.91$. In Figs.\ 7, 8 below we
shall explicitly show that the radii (\ref{rout},\ref{rside})
are reasonably good approximations for the actual inverse widths
of the correlation functions. The most important term
in (\ref{rout}) turns out to be $\beta_K^2 \, \langle \tilde{t}^2
\rangle$. Since $\beta_K^2 \simeq 0.27$  for the kaon pair, but
$\beta_K^2 \simeq 0.83$ for the pion pair, even if the
time variances $\langle \tilde{t}^2 \rangle$ of the emission functions
for kaon and pion pairs were roughly equal, 
the outward radius would turn out to be much smaller
in the kaon case. 

In fact, the time variance for the kaon emission function is in general
smaller than that of the pion source. In order to understand this, note
that the maximum of the emission function is typically located near
points of the freeze-out hypersurface where the particle pair velocity
$\beta_K$ matches the fluid velocity $\beta$, since this minimizes
the argument of the exponential term $\exp [-K \cdot u/T]$ in the
emission function. For a
larger particle mass, however, the emission function decreases faster
around this maximum, which in turn reduces its variance
as compared to the case of a smaller mass particle. 
Since this effect, however, also influences the spatial variance of
the emission function, we expect it to (roughly) cancel out in
the ratio $R_{\rm out}/R_{\rm side}$.
Thus, the suppression of 
$R_{\rm out}/R_{\rm side}$ for kaons as observed in Fig.\ 2 is mainly
due to the smaller kaon pair velocity.

On the other hand, following this argument we expect that
for increasing values of $K_\bot$, for which the
pair velocity for kaons is closer to the causal limit (and to that of pions),
$R_{\rm out}/R_{\rm side}$ should approach similar values 
as for pions. This is indeed seen to be the case in Fig.\ 3 where we plot 
$R_{\rm out}/R_{\rm side}$ for $K_{\bot}=700$ MeV and $\Delta T=0$.
As in the pion case, there is a strong
enhancement of about a factor of 2 over the corresponding
ideal gas case (thin lines) for all three freeze-out temperatures.
On the other hand, the dependence of $R_{\rm out}/R_{\rm side}$ 
on the freeze-out temperature is larger than at
$K_{\bot}=300$ MeV (cf. Fig.\ 2). The time--delay signal is strongest
for late freeze-out ($T_f \sim 0.7\, T_c$).

To study the influence of $K_\bot$ on the correlation
signal in more detail, we present in Fig.\ 4 the
ratio $R_{\rm out}/R_{\rm side}$ as a function of $K_{\bot}$
at fixed ${\epsilon}_{0}=18.75\,T_{c} s_{c}$, i.e., where the
time--delay signal is maximized, and for the case of a first order
transition in the EoS. For pions (thin lines) as well as for kaons
(thick lines) we observe a strong increase of the ratio with $K_\bot$. 
While for pions, $R_{\rm out}/R_{\rm side}$ is of order 3 
already at low momenta, the ratio for kaons increases only gradually
from around 1 at $K_\bot = 100$ MeV, on account of the above discussed
suppression due to the smaller kaon pair velocity. As expected, for
high momenta, $K_\bot \sim 1$ GeV, 
$R_{\rm out}/R_{\rm side}$ for kaons approaches values comparable 
to those for pions, since both pion and kaon pair velocities approach unity.

It is also seen from Fig.\ 4 that the observation made in
\cite{RisGySig} that the time--delay signal does not depend strongly
on the freeze-out temperature is not valid in general and was only due 
to the specific choice $K_\bot=300$ MeV in that work.
The sensitivity of the value of $R_{\rm out}/R_{\rm side}$ 
on the freeze-out temperature increases with $K_\bot$, 
such that $R_{\rm out}/R_{\rm side}$ is the larger the smaller
the freeze-out temperature is and may vary up to 20\% for
freeze-out between $T_f=0.7\, T_c$ and $T_f = T_c$.

In Fig.\ 5 we present the main result of the present
work: shown are surface plots of $R_{\rm out}/R_{\rm side}$ in the
$(\epsilon_0 - K_{\bot})$--plane.
The freeze-out temperature was fixed at $T_f=0.7\, T_c$.
Fig.\ 5 (a) shows the ratio of inverse widths for pions and $\Delta T=0$
(solid lines) versus the ideal gas case (dotted lines). In part (b)
the same is shown for $\Delta T=0.1\, T_c$, while (c) and (d)
show the corresponding results for kaons.
As one observes, the overall behaviour of $R_{\rm out}/R_{\rm side}$
is rather similar for pions, Figs.\ 5 (a,b), as for kaons, parts (c,d).
In accordance with Fig.\ 4 and for the (kinematic) reasons discussed above, 
the ratio at low $K_\bot$ is smaller for kaons than for pions. 
Also, the enhancement over the ideal gas case
is less pronounced in the case of a smooth transition,
$\Delta T=0.1\, T_c$, than for a first order phase transition, for reasons
discussed in detail in Ref.\ \cite{RisGySig}.

The most important property of Fig.\ 5 is to allow to identify
the range of initial energy densities and transverse pair momenta
required to maximize the time--delay signature. For our choice
of $\tau_0 = 0.1\, R_0 \simeq 0.5$ fm, 
the signal is maximized for initial energy 
densities around $\epsilon_0 \sim
20\, T_c s_c \simeq 15\, {\rm GeV\, fm}^{-3}$. For smaller (larger)
$\tau_0$, the respective values for $\epsilon_0$ increase (decrease),
as explained in \cite{RisGySig}. These values are
well above the soft region of the
EoS, cf.\ Fig.\ 1 (d). As discussed in detail in Ref.\ \cite{RisGySig}, this
is due to the strong longitudinal motion
in scaling hydrodynamics and opens the possibility to observe 
the time--delay effect at the RHIC collider. Moreover, Fig.\ 5 confirms
that, for sufficiently high $K_\bot$, the signal
is as strong for kaons as for pions. Kaon interferometry is, however,
advantageous from the experimental point of view, since kaon yields 
in general are less contaminated by resonance decays, while neutral kaon
correlations in particular are not subject to Coulomb distortions from
final state interactions.

In the following, we want to elucidate the behaviour of 
$R_{\rm out}/R_{\rm side}$ in greater detail by studying the inverse
widths separately as functions of $\epsilon_0$ and $K_{\bot}$. 
Fig.\ 6 shows (a,c) $R_{\rm out}$ and (b,d)
$R_{\rm side}$ both for (a,b) pions and (c,d) kaons for the 
case of a first order phase transition, $\Delta T=0$.
As one observes, for fixed $K_\bot$, $R_{\rm side}$ increases slowly
with $\epsilon_0$. This behaviour is naturally explained by the fact that
higher initial energy densities drive the transverse expansion of the 
system more strongly (cf.\ Ref.\ \cite{RisGySig} for the explicit hydrodynamic
solutions), which increases the transverse size of the source.
On the other hand, for fixed $\epsilon_0$, $R_{\rm side}$ slowly decreases
with increasing $K_\bot$.
To understand this quantitatively, recall that the
emission function (\ref{emiss}) has a maximum where $K \cdot u$ is
minimized. For larger pair momenta $K_\bot$ (larger pair 
velocities $\beta_K$) the
emission function $S(x,K) \sim \exp[ - K \cdot u/T]$ will 
drop faster in the region around the maximum, which in turn
makes the source appear smaller.

On the other hand, the $\epsilon_0$--dependence of $R_{\rm out}$
shows again clearly the time--delay signature of the QCD transition.
An unexpected feature is, however, that $R_{\rm out}$
{\em increases\/} with $K_\bot$ at fixed $\epsilon_0$, at least for the
case $\Delta T=0$ and for initial energy densities below about
$100\, T_c s_c$. According to the above argument, $R_{\rm out}$ is
expected to decrease as well, and the time--delay signature of
Fig.\ 5 solely due to the fact that the decrease of $R_{\rm out}$ is
{\em slower\/} than that of $R_{\rm side}$.

In order to understand why this is not the case for
$\Delta T=0$, let us first consider
a situation where this expectation is actually fulfilled.
Let us consider the hydrodynamic expansion in the case
of a smooth transition $\Delta T=0.1\, T_c$ and for an
initial energy density $\epsilon_0=18.75\, T_c s_c$ (cf.\ Fig.\ 9 (c)
of Ref.\ \cite{RisGySig}).
In Fig.\ 7 (a) we show the corresponding
isotherms in the $(t-r)$--plane (this figure is identical with
Fig.\ 9 (d) of \cite{RisGySig}), part (b)
shows the flow velocity $\beta$ along the freeze-out hypersurface, i.e.,
in our case the isotherm with $T=0.7\, T_c$ in Fig.\ 7 (a).
Since the mapping of the isotherm on time is unique we prefer
to plot $\beta$ as function of $t$ rather than as function of $r$, which,
considering the shape of the isotherm, is obviously not unique.

In Fig.\ 7 (c) we show the emission functions $S(x,K)$ for pions, integrated
over $\phi$ and $\eta$
(i.e., more precisely, the integrand of eq.\ (\ref{norm}) in
the appendix), as a function of time along the freeze-out isotherm 
for three different values of the average transverse pair momentum, 
$K_\bot = 300,\, 700,$ and $1000$ MeV.
As one expects, for larger pair momenta/velocities the maximum of the emission
function shifts to regions of higher
transverse velocity on the hypersurface, i.e., to earlier times. 
Also, in confirmation of the
above argument, the emission function becomes narrower, which reflects in 
a decreasing $R_{\rm out}^2$, as shown in Fig.\ 7 (e).
As expected, $R_{\rm side}^2$ decreases with $K_\bot$, too, although
this cannot be seen directly from
the presentation of the emission function in Fig.\ 7 (c).

Further understanding of this result can be gained by studying the
radii (\ref{rout},\ref{rside}). According to eq.\ (\ref{rout}),
$\hat{R}_{\rm out}^2$ consists of three different terms. The dashed line 
in Fig.\ 7 (e) shows the actual time variance $\langle \tilde{t}^2
\rangle$ of the emission function. In accordance with the above
arguments, and in agreement with Fig.\ 7 (c), 
this quantity decreases with $K_\bot$. For the dotted line,
the time variance was multiplied by $\beta_K^2$. For the light pions,
$\beta_K^2$ is close to unity already for moderate $K_\bot \sim
300$ MeV and the difference between these two curves rapidly vanishes
with increasing $K_\bot$. For the
long-dashed line the term $- 2 \, \beta_K \langle \, \tilde{x} \tilde{t} \,
\rangle$ was added. Obviously, the $x-t$--correlation of the emission
function is negative and about 50\% of the value of $\beta_K^2 \,
\langle \tilde{t}^2 \rangle$.
The dash-dotted line is the full result for $\hat{R}_{\rm out}^2$.
As one notices, the variance in $x$--direction is only a minor correction
to the other two terms. It is astonishing how well the approximate
relations (\ref{rout},\ref{rside}) reflect the qualitative {\em and\/}
quantitative behaviour of the inverse widths of the correlation
functions, although the shape of the
emission function is not Gaussian. (Note that 
all variances appearing in eqs.\ (\ref{rout},\ref{rside}) 
have been multiplied with a factor
1.44, to account for the difference in the definition of inverse widths
and radii, as discussed at the end of Section 3.)

In Figs.\ 7 (d,f) we show the corresponding results for kaons.
One observes that the emission function in the case $K_\bot
=300$ MeV is now somewhat narrower in time direction
than in the pionic case. The reason
is the larger kaon mass, or equivalently, the
smaller kaon pair velocity which suppresses
emission of such pairs from regions of the hypersurface with high
flow velocity, i.e., at early times, cf.\ Fig.\ 7 (b). (Remember that the
emission function is large in regions where pair and flow velocity are
equal.) This effect leads to considerably
smaller values for $R_{\rm out}$ and, as already discussed in
the context of Figs.\ 4 and 5, $R_{\rm out}/R_{\rm side}$.

For larger $K_\bot$, however, the size of the
region in time from which kaon pairs are
predo\-mi\-nant\-ly emitted does not decrease very much, as one confirms
via the time variance (dashed line in Fig.\ 7 (f)).
This at first sight astonishing result is 
due to the fact that the flow velocity remains large up to very late
times, cf.\ Fig.\ 7 (b), 
such that kaon pairs with high velocity are not only emitted
at early times (where the flow velocity is high anyway), but also
up to late times before the origin cools below $T=0.7\, T_c$.
The strong increase in the kaon pair velocity with $K_\bot$ 
leads then to a corresponding
{\em increase\/} in $R_{\rm out}$ and $\hat{R}_{\rm out}$, cf.\ Fig.\
7 (f).Note that due to the larger kaon mass, the difference between
dashed and dotted curves is much bigger than for the pion case.

In Fig.\ 8 we present the corresponding results for the case
$\Delta T=0$ which serves to explain the behaviour observed in
Fig.\ 6. In this case, the outward radii increase with $K_\bot$
{\em both\/} for kaons {\em and\/} pions. 
The reason is that the emission functions,
besides the shift of the maximum
to earlier times, do not exhibit a decrease of the emission probability
at late times, resulting in an overall {\em increase\/} in the duration
of particle emission and, correspondingly, in the outward radii.
From the above it is clear that due to the larger kaon mass, the increase of
the outward radius with $K_\bot$ is much stronger for kaons than for pions.

The reason for the increase in the width of the emission function
is the behaviour of the flow velocity, Fig.\ 8 (b). 
Even up to times where the freeze-out
reaches the origin, $\beta$ stays comparatively large, enhancing the
emission probability for particle pairs with high velocity.
The physical reason for this high transverse velocity is a rarefaction
shock wave (cf.\ Fig.\ 9 (a) in \cite{RisGySig}), which travels
inwards and expells matter with a constant flow velocity, a fact that
is confirmed by the plateau observed in Fig.\ 8 (b) for times
between $3$ and $6\, R_0$. Note again the excellent agreement between
approximate radii, $\hat{R}_{\rm out,\, side}$,
and the inverse widths of the correlation
functions, $R_{\rm out,\, side}$, 
in Figs.\ 8 (e,f), although the emission functions are
not even remotely resembling Gaussian distributions.
In fact, for (\ref{rout},\ref{rside}) to be good approximations of
the inverse widths it is sufficient that the {\em correlation functions\/}
resemble Gaussians, rather than the emission functions themselves.
In most (realistic) situations, this happens to be the case \footnote{We
thank U.\ Heinz for pointing this out.}.

We finally mention that the high flow velocity in the cases
of interest ensures that the negative contributions to the
spectra and the correlation functions are negligible, 
as indicated in Subsection 2.3.
Quantitatively, they are smaller than 1\% (of the total particle number
crossing the freeze-out hypersurface) for
the smallest transverse momenta considered, $K_\bot \sim 100$ MeV.
For larger transverse momenta, they are even smaller.

\section{Summary and Conclusions}

In this work we have investigated the time--delay effect
of the QCD transition on two--pion and two--kaon correlations
in ultrarelativistic heavy-ion collisions in the framework of ideal
hydrodynamics with transverse cylindrical and boost-invariant longitudinal
symmetry. Our study is an extension of the work in \cite{RisGySig},
where this time--delay signal, i.e., an enhanced ratio 
$R_{\rm out}/R_{\rm side}$, was investigated for
pion interferometry with fixed average transverse pair
momentum $K_\bot=300$ MeV.
Our main interest in the present work was the transverse momentum
dependence of the proposed signal, and the question whether
it can be also seen in two--kaon interferometry.
Kaon interferometry seems preferable for two reasons. First, 
effects of resonance decays are small
(only the shorter-lived $K^{\star}$
resonances influence the final kaon spectra, at most a (small) 10\%--effect
at temperatures corresponding to freeze-out \cite{resonance}).
For pion interferometry, however, the decay of long-lived
resonances plays an important role and virtually increases
the (average) source size and lifetime of the system
\cite{padula}. This results in considerable modifications of the inverse
width of the two--particle correlation functions of pions
\cite{resonance,Marburg,Schlei}.
Second, neutral kaon correlation functions would not be influenced by
final state Coulomb interactions at small
relative momenta (where the correlation signal is large).

We presented a systematic study of the inverse widths $R_{\rm out}$,
$R_{\rm side}$, and their ratio for two--pion and two--kaon correlation 
functions at midrapidity, as a function of the
initial energy density $\epsilon_0$, the average transverse momentum
of the particle pair $K_\bot$, and for different equations of state. 
Studying the emission functions and
their variances for selected cases, we explained in detail the momentum
dependence of the inverse width in out--direction.
We confirmed that the prolongation of the lifetime of the system
due to the QCD transition is observable in an enhanced
ratio $R_{\rm out}/R_{\rm side}$ for kaons as well as for pions.
The proposed signal is maximized for initial energy densities 
expected to be reached in central Au+Au--collisions at the RHIC collider,
$\epsilon_0 \sim 10 - 20 \, {\rm GeV\, fm}^{-3}$ (for 
realistic \cite{hotglue} values of the initial thermalization time
scale, $\tau_0 \sim 0.5$ fm),
and for pairs with high average transverse momentum, $K_\bot \sim
1$ GeV. This opens the
interesting possibility to observe the time--delay signal of the
QCD transition with kaon interferometry in the STAR and
PHENIX experiments at RHIC.
\\ ~~ \\
\noindent
{\bf Acknowledgements}
\\ ~~ \\
We would like to thank M.\ Gyulassy, T.\ Hallman, U.\ Heinz, B.\ Schlei,
R.\ Pisarski, and W.\ Zajc for stimulating discussions. D.H.R.\ thanks in
particular U.\ Heinz for encouraging this study, for a critical
reading of the manuscript, and for invaluable
help in interpreting the results. Special thanks go to M.\ Gyulassy
for his continued support and encouragement throughout this work.
D.H.R.\ gratefully acknowledges the hospitality of 
the Institute for Theoretical Physics of the
University Frankfurt. The work of D.H.R.\ has been supported by the
Director, Office of Energy
Research, Division of Nuclear Physics of the Office of 
High Energy and Nuclear Physics of the
U.S.\ Department of Energy under contract nos.\ DE-FG-02-93ER-40764
and DE-FG02-96ER40945.
\newpage
\appendix
\section*{Appendix}
In this appendix we collect the explicit expressions necessary to
evaluate the variances of the emission function $S(x,K)$ appearing
in eqs.\ (\ref{rout},\ref{rside}). We start with noting that
\begin{eqnarray}
\langle \tilde{x}^2 \rangle & = & \langle x^2 \rangle - \langle x \rangle^2\,\,
, \nonumber \\
\langle \tilde{t}^2 \rangle & = & \langle t^2 \rangle - \langle t \rangle^2\,\,
, \\
\langle \, \tilde{x} \tilde {t} \, \rangle & = & \langle x t \rangle - 
\langle x \rangle \, \langle t \rangle \, , \nonumber \\
\langle \tilde{y}^2 \rangle & = & \langle y^2 \rangle - \langle y \rangle^2\,\,
, \nonumber
\end{eqnarray}
while due to the $\delta$--function in eq.\ (\ref{emiss}), all
space--time variables are taken along the freeze-out hypersurface $\Sigma$,
given by the 4--vector
\begin{equation}
\Sigma^{\mu} = \left (\tau_f(\zeta)\, \cosh \eta, \, r_f(\zeta)\, \cos \phi,
\, r_f (\zeta) \, \sin \phi, \, \tau_f(\zeta)\, \sinh \eta \right)\,\, ,
\end{equation}
with $r_f(\zeta),\, \tau_f(\zeta) \equiv t_f(\zeta,\eta=0)$, given
by the isotherms in Figs.\ 7, 8 (a).
Here we adhered to the notation of Ref.\ \cite{RisGySig}, i.e.,
the hypersurface $\Sigma$ is parametrized by the three variables
$\zeta,\, \eta,$ and $\phi$, where $\zeta$ runs from 0 to 1 along the
hypersurface in the $(t-r)$--plane at $z=0$, $\eta$ from $-\infty$ to
$\infty$ along the hypersurface in the $(t-z)$--plane ($\eta$ is identical
to the space--time rapidity ${\rm Artanh} [z/t]$) and $\phi$ from 0 to
$2 \pi$ in the transverse $(x-y)$--plane. Due to the transverse
cylindrical and boost-invariant longitudinal symmetry
of the problem, the parametrization of the hypersurface in terms of
$\phi$ and $\eta$ is trivial. 
Note that for $z=\eta=0$, $\tau \equiv \sqrt{t^2 - z^2} \equiv t$.

Using the definition of the average $\langle\, \cdot \,\rangle$ in eq.\
(\ref{apprC2}), we find after employing (cf.\ eq.\ (B.3) of \cite{RisGySig}) 
\begin{equation} \label{dsigmadotK}
{\rm d} \Sigma \cdot K = \left(- E_K \cosh \eta \, \frac{{\rm d}
r_f}{{\rm d} \zeta} + K_\bot  \cos \phi \,\, \frac{{\rm d}
\tau_f}{{\rm d} \zeta} \right) r_f(\zeta) \, \tau_f (\zeta) \, {\rm d}
\zeta \, {\rm d} \eta\, {\rm d} \phi\,\, ,
\end{equation}
where we used $K^0 \simeq E_K$, ${\bf K} = (K_\bot,0,0)$,
and (cf.\ eq.\ (B.4) of \cite{RisGySig})
\begin{equation} \label{umu}
K\cdot u  = E_K \cosh \eta\, \cosh \eta_r - K_\bot \sinh \eta_r \cos 
\phi\,\, ,
\end{equation}
where $\eta_r \equiv {\rm Artanh}\, \beta$ ($\beta$ is the radial
component of the transverse velocity at $z=0$), and after
integrating over $\eta, \, \phi$ the final result:
\begin{eqnarray}
\langle x \rangle & = & {\cal N}^{-1} \int_0^1 {\rm d}\zeta\,\, r_f^2 \tau_f
\left\{ - E_K\, \frac{{\rm d}r_f}{{\rm d}\zeta}\, K_1(a)\, I_1(b)
        + K_\bot\, \frac{{\rm d}\tau_f}{{\rm d}\zeta}\, K_0(a)\, \left[
        I_0(b)-\frac{I_1(b)}{b} \right] \right\} \,\, , \\
\langle y \rangle & = & 0 \,\, , \\
\langle t \rangle & = & {\cal N}^{-1} \int_0^1 {\rm d}\zeta\,\, r_f \tau_f^2
\left\{ - E_K\, \frac{{\rm d}r_f}{{\rm d}\zeta}\, \left[ K_0(a) + 
       \frac{K_1(a)}{a} \right]  I_0(b)
      + K_\bot\, \frac{{\rm d}\tau_f}{{\rm d}\zeta}\, K_1(a)\, I_1(b)
       \right\} , \\
\langle x^2 \rangle & = & {\cal N}^{-1} \int_0^1 {\rm d}\zeta\,\, r_f^3 \tau_f
\left\{ - E_K\, \frac{{\rm d}r_f}{{\rm d}\zeta}\, K_1(a)\, \left[
        I_0(b) - \frac{I_1(b)}{b} \right] \right. \nonumber \\
   &  & \left.\hspace*{3.2cm} + K_\bot\, 
       \frac{{\rm d}\tau_f}{{\rm d}\zeta}\, K_0(a)\, \left[
     -\frac{I_0(b)}{b} +\left(1+\frac{2}{b^2}\right) \, I_1(b) \right] 
      \right\} \,\, , \\
\langle y^2 \rangle & = & {\cal N}^{-1} \int_0^1 {\rm d}\zeta\,\, r_f^3 \tau_f
\left\{ - E_K\, \frac{{\rm d}r_f}{{\rm d}\zeta}\, K_1(a)\, \frac{I_1(b)}{b} 
   \right.  \nonumber \\
 &   & \left. \hspace*{3.2cm}  + K_\bot\, 
      \frac{{\rm d}\tau_f}{{\rm d}\zeta}\, K_0(a)\, \left[
        \frac{I_0(b)}{b}- \frac{2\, I_1(b)}{b^2} \right] \right\} \,\, , \\
\langle t^2 \rangle & = & {\cal N}^{-1} \int_0^1 {\rm d}\zeta\,\, r_f \tau_f^3
\left\{ - E_K\, \frac{{\rm d}r_f}{{\rm d}\zeta}\, \left[
       \frac{ K_0(a)}{a} + \left(1+\frac{2}{a^2}\right) K_1(a) 
        \right]\, I_0(b) \right. \nonumber \\
&   & \left. \hspace*{3.2cm} + K_\bot\, 
       \frac{{\rm d}\tau_f}{{\rm d}\zeta}\,  \left[
        K_0(a)+\frac{K_1(a)}{a} \right] \, I_1(b) \right\} \,\, , \\
\langle x t \rangle & = & {\cal N}^{-1} \int_0^1 {\rm d}\zeta\,\, r_f^2 
        \tau_f^2 \left\{ - E_K\, \frac{{\rm d}r_f}{{\rm d}\zeta}\, \left[
        K_0(a) + \frac{K_1(a)}{a} \right]\, I_1(b) \right. \nonumber \\
 &   & \left. \hspace*{3.2cm} + K_\bot\, 
        \frac{{\rm d}\tau_f}{{\rm d}\zeta}\, K_1(a) \, \left[
        I_0(b)-\frac{I_1(b)}{b} \right] \right\} \,\, , \\
{\cal N} & = & \int_0^1 {\rm d}\zeta\,\, r_f \tau_f
\left\{ - E_K\, \frac{{\rm d}r_f}{{\rm d}\zeta}\, K_1(a) \, I_0(b)
      + K_\bot\, \frac{{\rm d}\tau_f}{{\rm d}\zeta}\, K_0(a)\, I_1(b)
       \right\} \,\, , \label{norm}
\end{eqnarray}
where $a\equiv E_K \cosh \eta_r/T$, $b\equiv K_\bot \sinh \eta_r/T$.

\newpage
\noindent

{\bf Figure Captions:}
\\ ~~ \\
{\bf Fig.\ 1:}
(a) The entropy density divided by $T^3$
(in units of $s_c/T_c^3$), (b) the energy density divided by 
$T^4$ (in units of $T_c\, s_c/ T_c^4$) as functions of temperature
(in units of $T_c$),
(c) the pressure (in units of $T_c\, s_c$), (d) the square of
the velocity of sound as functions of energy density
(in units of $T_c\, s_c$). The solid lines
correspond to $\Delta T=0$, the dotted curves to
$\Delta T=0.1\, T_c$. Quantities for the ideal gas EoS
(with $d_{\rm H}$ degrees of freedom) are represented by dashed lines.
The ratio of degrees of freedom in the QGP to those in the hadronic phase is
$d_{\rm Q}/d_{\rm H}=37/3$. The critical enthalpy density
is $T_c\, s_c \simeq  0.75 {\rm GeV\, fm}^{-3}$ for the case $d_{\rm Q}=37,\, 
d_{\rm H}=3$.
\\ ~~ \\
{\bf Fig.\ 2:}
$R_{\rm out}/R_{\rm side}$ as a function of the initial energy density
$\epsilon_0$ for the Bjorken cylinder expansion ($\tau_0 = 0.1\, R_0$)
with a first order phase transition ($\Delta T=0$).
Thin lines correspond to pion pairs ($m_\pi=138$ MeV), 
thick lines to kaons ($m_K=497$ MeV). The pairs are at midrapidity
and have average transverse momentum $K_{\bot}=300$ MeV.
The results are shown for freeze-out
temperatures $T_{f}=0.7\,T_{c}$ (full lines), 
$T_{f}=0.9\,T_{c}$ (dotted lines), and $T_{f}=T_{c}$ (dashed lines).
\\ ~~ \\
{\bf Fig.\ 3:}
$R_{\rm out}/R_{\rm side}$ for kaons 
as a function of the initial energy density
${\epsilon}_{0}$ at $K_{\bot}=700$ MeV. The
ideal gas case (thin lines) is compared to that of a
first order phase transition (thick lines,
$\Delta T=0$). Curves correspond to different
freeze-out temperatures as in Fig.\ 2.
\\ ~~ \\
{\bf Fig.\ 4:}
The dependence of $R_{\rm out}/R_{\rm side}$ on the
average transverse momentum $K_{\bot}$ of the particle pair, for
fixed energy density ${\epsilon}_{0}=18.75\,T_{c} s_{c}$.
Thin lines correspond to pions, thick lines to kaons.
Different curves correspond to different freeze-out temperatures,
as in Fig.\ 2.
\\ ~~ \\
{\bf Fig.\ 5:} 
(a) $R_{\rm out}/R_{\rm side}$ as a 
function of $\epsilon_0$ and $K_\bot$ for pions at midrapidity
and freeze-out temperature $T_{f}=0.7\,T_{c}$,
for the case of a first order phase transition ($\Delta T=0$, full lines)
versus the ideal gas case (dotted).
(b) as in (a), but for $\Delta T=0.1\, T_c$. (c,d) as in (a,b), but for
kaons.
\\ ~~ \\
{\bf Fig.\ 6:} 
(a) Inverse width of the pion correlation function in out--direction
$R_{\rm out}$ (in units of the initial transverse radius $R_{0}$)
at midrapidity and for $\Delta T=0$,
as a function of $\epsilon_0$ and $K_\bot$.
(b) as in (a), but for $R_{\rm side}$. (c,d) as in (a,b), but for
kaons.
\\ ~~ \\
{\bf Fig.\ 7:}
(a) Isotherms in the $(t-r)$--plane (at $z=0$) for the hydrodynamic expansion
of a Bjorken cylinder with $\epsilon_0 = 18.75\, T_c s_c$, $\tau_0 = 0.1\,
R_0$, and
for a smooth transition $\Delta T=0.1\, T_c$ in the EoS. Labels correspond
to temperatures in units of $T_c$. The freeze-out isotherm ($T_f = 0.7\, T_c$)
is the thick line.
(b) The flow velocity $\beta$ as a function of time
along the freeze-out isotherm of part (a).
(c) The emission function for pions (in arbitrary units), 
integrated over $\eta$ and $\phi$
(and smoothed over approximately 100 time steps of the hydrodynamical
evolution), along the
freeze-out isotherm of part (a), for $K_\bot = 300$ MeV (solid line),
$K_\bot = 700$ MeV (dotted line), and $K_\bot = 1000$ MeV (long-dashed
line). (d) as in (c), but for kaons. (e) $R_{\rm out}^2$ and $R_{\rm side}^2$
for pions as a function of $K_\bot$ (solid lines), and the approximate
radii $\hat{R}_{\rm out}^2$ and $\hat{R}_{\rm side}^2$ (dash-dotted lines).
The dashed line is the variance in time only, $\langle \tilde{t}^2 \rangle$, 
while for the dotted line, this quantity was multiplied with
$\beta_K^2$. For the long-dashed line, the $x-t$--correlation 
$-2\, \beta_K \langle \,\tilde{x}\tilde{t}\, \rangle$ has been added.
(f) as in (e), but for kaons.
\\ ~~ \\
{\bf Fig.\ 8:}
As in Fig.\ 7, but for $\Delta T= 0$.


\newpage
\begin{thebibliography}{99}

\bibitem{Lattice}
F.\ Karsch, Nucl.\ Phys.\ B (Proc.\ Suppl.) 34 (1994) 63,
Phys.\ Rev.\ D 49 (1994) 3791, \\
E.\ Laermann, Nucl.\ Phys.\ A 610 (1996) 1c.

\bibitem{muller}
B.\ M\"uller, Rep.\ Prog.\ Phys.\ 58 (1995) 611.

\bibitem{harris}
J.W.\ Harris and B.\ M\"uller, Annu.\ Rev.\ Nucl.\ Part.\ Sci.\ 46 (1996) 71.

\bibitem{strottman}
R.B.\ Clare and D.D.\ Strottman, Phys.\ Rep.\ 141 (1986) 177.

\bibitem{Shuryak}
C.M.\ Hung and E.V.\ Shuryak, Phys.\ Rev.\ Lett.\ 75 (1995) 4003.

\bibitem{RisBerM95}
D.H.\ Rischke, S.\ Bernard, J.A.\ Maruhn, Nucl.\ Phys.\ A 595 (1995) 346.

\bibitem{RisGy96}
D.H.\ Rischke and M.\ Gyulassy, Nucl.\ Phys.\ A 597 (1996) 701.

\bibitem{Csernai}
L.V.\ Bravina, N.S.\ Amelin, L.P.\ Csernai, P.\ Levai,
D.\ Strottman, Nucl.\ Phys.\ A 566 (1994) 461c.

\bibitem{RisPuretal}
D.H.\ Rischke, Y.\ P\"urs\"un, J.A.\ Maruhn,
H.\ St\"ocker, W.\ Greiner, Heavy Ion Phys.\ 1 (1995) 309.

\bibitem{RisGySig}
D.H.\ Rischke and M.\ Gyulassy, Nucl.\ Phys.\ A 608 (1996) 479.

\bibitem{Pratt}
S.\ Pratt, Phys.\ Rev.\ C 49 (1994) 2722, Phys.\ Rev.\
D 33 (1986) 1314.

\bibitem{Bertsch}
G.\ Bertsch, M.\ Gong, M.\ Tohyama, Phys.\ Rev.\ C 37 (1988)
1896,\\
G.\ Bertsch, Nucl.\ Phys.\ A 498 (1989) 173c.

\bibitem{resonance}
J.\ Bolz, U.\ Ornik, M.\ Pl\"umer, B.R.\ Schlei, R.M.\ Weiner,
Phys.\ Rev.\ D 47 (1993) 3860,\\
B.\ Schlei, U.\ Ornik, M.\ Pl\"umer, D.\ Strottman, R.M.\ Weiner,
Phys.\ Lett.\ B 376 (1996) 212.

\bibitem{heinz}
U.A.\ Wiedemann, P.\ Scotto, U.\ Heinz, Phys.\ Rev.\ C 53 (1996)
918, \\
U.\ Heinz, lectures given at the NATO ASI on ``Correlations
and Clustering Phenomena in Subatomic Physics'', Dronten, Netherlands,
August 4 -- 18, 1996, LANL preprint server nucl-th/9609029.

\bibitem{hotglue}
E.\ Shuryak, Phys.\ Rev.\ Lett.\ 68 (1992) 3270, \\
K.J.\ Eskola and M.\ Gyulassy, Phys.\ Rev.\ C 47 (1993) 2329.

\bibitem{LanLif}
L.D.\ Landau and E.M.\ Lifshitz, ``Fluid mechanics''
(Pergamon, New York, 1959).

\bibitem{Bjorken}
J.D.\ Bjorken, Phys.\ Rev.\ D 27 (1983) 140,\\
G.\ Baym, B.L.\ Friman, J.P.\ Blaizot, 
M.\ Soyeur, W.\ Czy\.{z}, Nucl.\ Phys.\ A 407 (1983) 541.

\bibitem{MGDHRBZ}
M.\ Gyulassy, D.H.\ Rischke, B.\ Zhang, Nucl.\ Phys.\ A 613 (1997) 397.

\bibitem{Sod}
G.A.\ Sod, J.\ Fluid Mech.\ 83 (1977) 785.

\bibitem{Schneider}
V.\ Schneider et al., J.\ Comput.\ Phys.\ 105 (1993) 92.

\bibitem{RisPurM95}
D.H.\ Rischke, Y.\ P\"urs\"un, J.A.\ Maruhn,
Nucl.\ Phys.\ A 595 (1995) 383.

\bibitem{Para}
J.P.\ Blaizot and J.Y.\ Ollitrault, Phys.\ Rev.\ D 36 (1987) 916.

\bibitem{bag}
A.\ Chodos, R.L.\ Jaffe, K.\ Johnson,
C.B.\ Thorn, V.F.\ Weisskopf, Phys.\ Rev.\ D 9 (1974) 3471.

\bibitem{carsten}
C.\ Greiner and C.\ Spieles, private communication (preprint in
preparation).

\bibitem{bugaev}
K.A.\ Bugaev, Nucl.\ Phys.\ A 606 (1996) 559.

\bibitem{jjn}
J.J.\ Neumann, B.\ Lavrenchuk, G.\ Fai, LANL preprint server
nucl-th/9612020.

\bibitem{CooFry74}
F.\ Cooper and G.\ Frye, Phys.\ Rev.\ D 10 (1974) 186, \\
F.\ Cooper, G.\ Frye, E.\ Schonberg, Phys.\ Rev.\ D 11 (1975) 192.

\bibitem{BerMGRis96}
S.\ Bernard, J.A.\ Maruhn, W.\ Greiner, D.H.\ Rischke,
Nucl.\ Phys.\ A 605 (1996) 566.

\bibitem{sinyukov}
Yu.M.\ Sinyukov, Z.\ Phys.\ C 43 (1989) 401.

\bibitem{Sinyukov}
Yu.M.\ Sinyukov, Nucl.\ Phys.\ A 498 (1989) 151c.

\bibitem{Marburg}
B.R.\ Schlei, U.\ Ornik, M.\ Pl\"umer, R.M.\ Weiner, Phys.\ Lett.\
B 293 (1992) 275.

\bibitem{YKP}
F.\ Yano and S.\ Koonin, Phys.\ Lett.\ B 78 (1978) 556,\\
M.I.\ Podgoretskii, Sov.\ J.\ Nucl.\ Phys.\ 37 (1983) 272,\\
S.\ Chapman, J.R.\ Nix, U.\ Heinz, Phys.\ Rev.\ C 52 (1995) 2694.

\bibitem{dinesh}
J.\ Alam, D.K.\ Srivastava, B.\ Sinha, D.N.\ Basu,
Phys.\ Rev.\ D 48 (1993) 1117.

\bibitem{padula}
S.S.\ Padula and M.\ Gyulassy, Nucl.\ Phys.\ A 544 (1992) 537c.

\bibitem{Schlei}
B.R.\ Schlei, Los Alamos preprint LA-UR-96-1614, LANL preprint server
nucl-th/9605016, \\
U.A.\ Wiedemann and U.\ Heinz, LANL preprint server nucl-th/9611031.
\end{thebibliography}
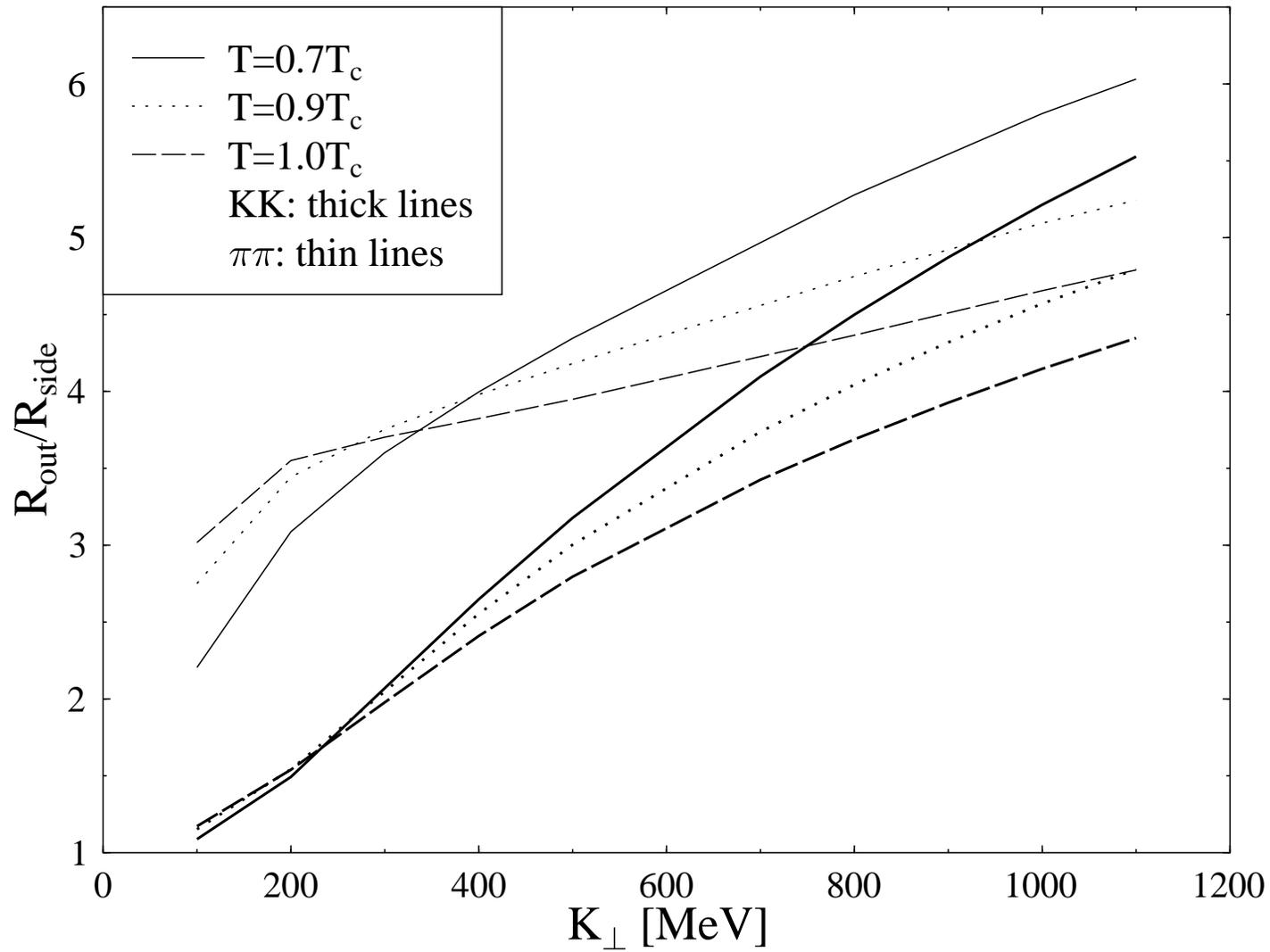
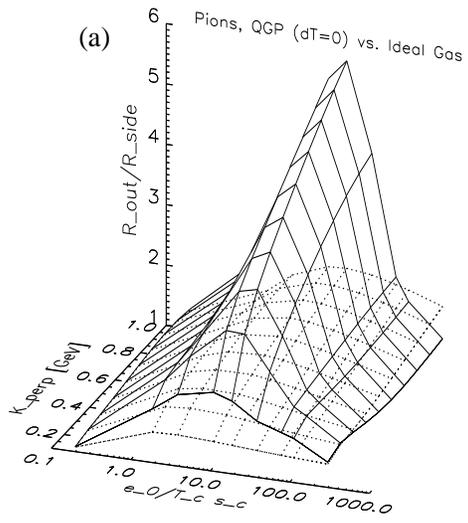
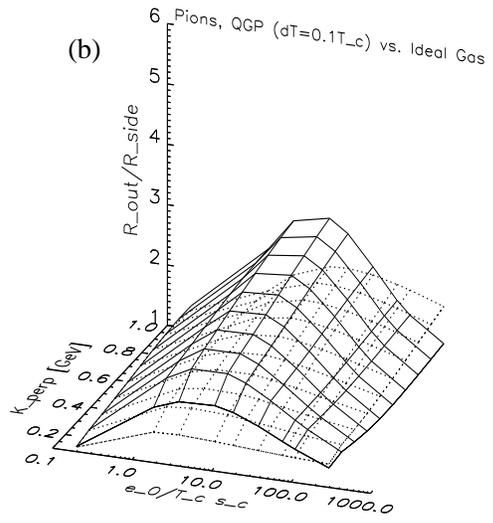
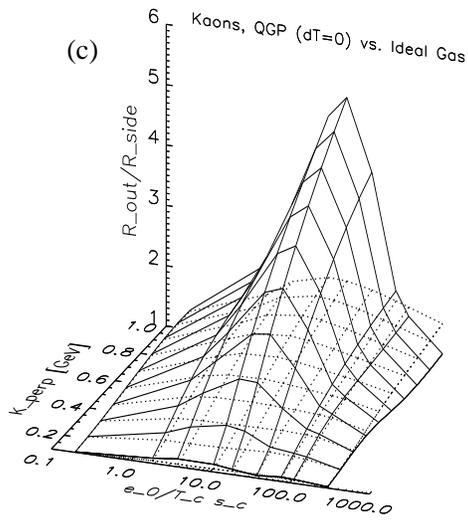
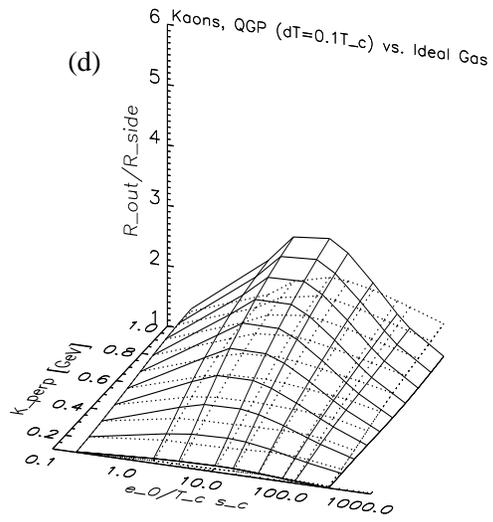
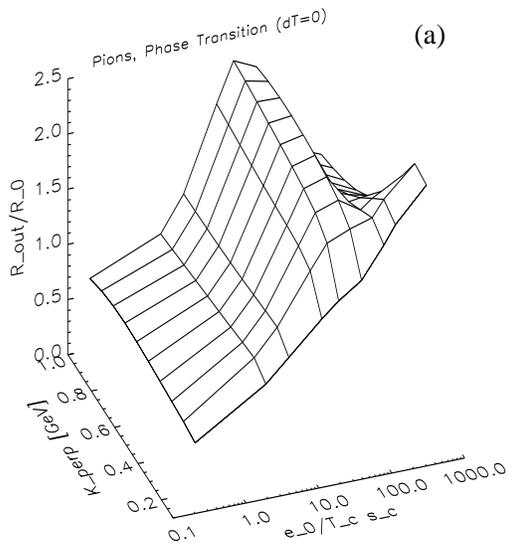
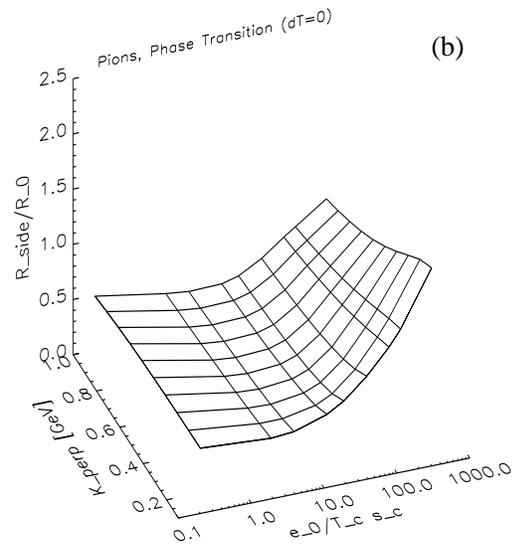
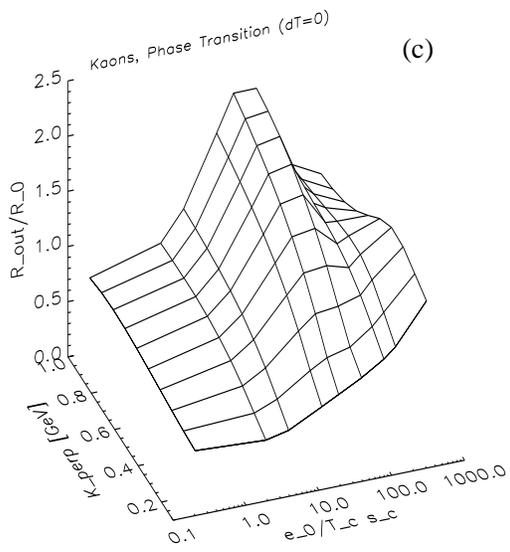
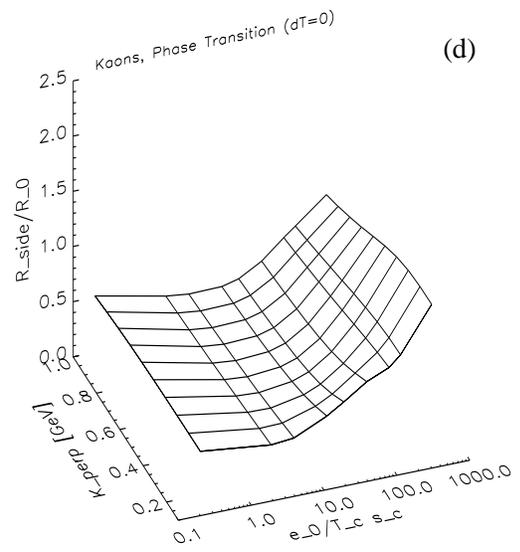
\end{document}